\documentstyle[12pt,equation]{article}

\begin{document}

\newcommand{\as}{\mbox{$\alpha_S$}}
\newcommand{\msqav}{\mbox{$\langle m_{\tilde{q}} \rangle$}}
\newcommand{\be}{\begin{equation}}
\newcommand{\ee}{\end{equation}}
\newcommand{\een}{\end{subequations}}
\newcommand{\ben}{\begin{subequations}}
\newcommand{\beq}{\begin{eqalignno}}
\newcommand{\eeq}{\end{eqalignno}}
\newcommand{\tanb}{\mbox{$\tan \beta$}}
\newcommand{\eplem}{\mbox{$e^+e^-$}}
\newcommand{\rs}{\mbox{$\sqrt{s}$}}
\newcommand{\non}{\nonumber}
\newcommand{\nn}{\noindent}
\newcommand{\mt}{m_q}
\newcommand{\msq}{\mbox{$m_{\tilde{q}}$}}
\newcommand{\msi}{\mbox{$m_{\tilde{q}_i}$}}
\newcommand{\msj}{\mbox{$m_{\tilde{q}_j}$}}
\newcommand{\mso}{\mbox{$m_{\tilde{q}_1}$}}
\newcommand{\mst}{\mbox{$m_{\tilde{q}_2}$}}
\newcommand{\mg}{\mbox{$M_{\tilde{g}}$}}
\newcommand{\is}{2I_q^{3L}}
\newcommand{\heta}{\tilde{\theta}}
\newcommand{\bbbar}{\mbox{$b \bar{b}$}}
\newcommand{\ttbar}{\mbox{$t \bar{t}$}}
\newcommand{\ccbar}{\mbox{$c \bar{c}$}}
\newcommand{\qqbar}{\mbox{$q \bar{q}$}}
\newcommand{\bq}{\beta_q}
\newcommand{\ct}{\cos^2\theta}
\newcommand{\st}{\sin^2\theta}
\renewcommand{\thefootnote}{\fnsymbol{footnote} }

\pagestyle{empty}

\begin{flushright}
MAD/PH/754 \\
UdeM-LPN--93--142 \\
OCIP/C--93--4 \\
April 1993 \\
\end{flushright}

\vspace*{1cm}
\begin{center}
{\large{\bf Supersymmetric QCD corrections to quark pair production
\\ \vspace*{3mm} in e$^+$e$^-$ annihilation}}

\vspace*{1.2cm}

A.~DJOUADI$^1\footnote{NSERC fellow}$, M.~DREES$^2$\footnote{Heisenberg fellow}
and H.~K\"ONIG$^3$

\vspace*{1cm}

\rm{$^1$ Lab. Phys. Nucl., Universit\'e de Montr\'eal, Case 6128A, H3C
3J7 Montr\'eal PQ, Canada}

\vspace*{.2cm}

\rm{$^2$ Physics Department, University of Wisconsin, Madison, WI 53706,
USA}

\vspace*{.2cm}

\rm{$^3$ Department  of Physics, Carleton University, Ottawa K1S 5B6, Canada}

\end{center}
\vspace*{1.5cm}

\begin{abstract}

\nn We calculate supersymmetric QCD corrections (squark/gluino loops) to quark
pair production in \eplem\ annihilation, allowing for mixing between left-- and
right--handed squarks and taking into account the effects of nonzero quark
masses. Corrections to the $Z$ boson partial widths are generally small and
positive, except in the case of large $\tilde{b}$ squark mixing, where they
become negative. At high--energy $e^+e^-$ colliders, larger corrections to the
total cross sections are possible. Corrections to forward--backward asymmetries
are negligible except possibly for top quarks, where they are sensitive to
$\tilde{t}$ squark mixing. We also comment on the possibility that the gluino
mass is only a few GeV.

\end{abstract}

\newpage

\pagestyle{plain}
\renewcommand{\thefootnote}{\arabic{footnote} }
\setcounter{footnote}{0}

\section*{1.~Introduction}

\nn The introduction of Supersymmetry \cite{1} (SUSY) is one of the most
attractive extensions of the Standard Model (SM). It not only stabilizes
\cite{2} the huge hierarchy between the weak scale and the Grand Unification
or Planck scale against radiative corrections; if SUSY is broken at a
sufficiently large scale, as is the case, e.g., in Supergravity (SUGRA)
models \cite{3}, it might allow to understand the origin of the hierarchy in
terms of radiative gauge symmetry breaking \cite{4}. Moreover, SUSY models
offer a natural solution of the cosmological Dark Matter problem \cite{5},
and allow for a consistent Grand Unification of all known gauge couplings,
in contrast to the nonsupersymmetric SM \cite{6}. All these attractive
features are already present in the minimal supersymmetric extension of the
SM, the MSSM, to which we will stick in this article. \\

\nn Unfortunately no direct signal for the production of superparticles has
yet been observed; experimental searches so far have only resulted in lower
bounds on sparticle masses, the most stringent ones coming from LEP \cite{7}
and the Tevatron \cite{8}. It is therefore tempting to look for SUSY through
precision measurements, where quantum corrections involving superparticles
might alter SM predictions. The potentially largest corrections can be expected
from corrections involving strong interactions, i.e. from squark and gluino
loops. Given the inherent uncertainties of cross section calculations as well
as measurements at hadron colliders, the most promising (and also the simplest)
process where such corrections can be probed is quark pair production in
\eplem\ annihilation\footnote{Squark and gluino loops also contribute to rare
$K$ and $B$ meson decays and oscillations. However, these corrections always
involve flavour--changing couplings, which in the MSSM are induced only through
weak interactions. As a result, in the MSSM supersymmetric QCD loops in $K$ and
$B$ meson physics are actually smaller \cite{9} than loops involving
electroweak gauginos or Higgs bosons.}. \\

\nn In this paper, we calculate the supersymmetric QCD corrections to quark
pair production in \eplem\ annihilation, allowing for mixing between left-- and
right--handed squarks and taking into account the effects of nonzero quark
masses. At LEP1 energies, we find that these corrections are small and positive
for $Z$ decays into light quarks; however for $b \bar{b}$ final states, mixing
in the $\tilde{b}$ squark sector can affect the correction to the cross
section, and can even change its sign. In the case of top quark pair production
at high--energy $e^+e^-$ colliders, the effect of mixing in the $\tilde{t}$
squark sector on the total cross section is less significant, since the
dominant photon exchange contribution is not sensitive to it. The correction to
the top forward--backward asymmetry does depend on the details of $\tilde{t}$
squark mixing but unfortunately the correction is always very small, and will
therefore be difficult to measure. \\

\nn Supersymmetric QCD corrections to quark pair production in \eplem\
annihilation were first discussed in Ref.~\cite{R9} for LEP1 energies in the
approximation of negligible quark masses and squark mixing and of equal masses
of the superpartners of left-- and right--handed quarks. In Ref.~\cite{R10} the
effect of squark mixing has been included at LEP1 energies and found to be
small. However, in that paper only corrections to the $Z$--quark couplings
present in the SM at tree level are considered, while we compute all CP
conserving form factors for both the $Z$ boson and the photon (the latter are
needed for c.m.~energies away from the $Z$ resonance).
In the limit of zero quark mass and squark mixing, our results for the total
cross section fully agree with Ref.\cite{R9} both numerically and analytically;
we also find general numerical agreement with Ref.~\cite{R10}. Finally, we also
compute corrections to the forward--backward asymmetry, while the previous
papers \cite{R9,R10} focussed on corrections to total rates. \\

\nn The rest of this paper is organized as follows. In sec. 2 we set up the
formalism and present our analytical results for the corrections to the most
general set of CP conserving $\gamma \qqbar$ and $Z \qqbar$ couplings. In sec.
3 we show numerical examples both for LEP1 and for a future high--energy
$e^+e^-$ linear collider operating at $\rs = 500$ GeV. Sec. 4 contains a
summary and some conclusions. For the convenience of the reader explicit
expressions for the scalar 2-- and 3--point functions appearing in our results
are listed in the Appendix.

\section*{2. Formalism}

\nn The most general $Z\qqbar$ and $\gamma \qqbar$ vertices
compatible with CP invariance can be written as
\begin{eqnarray} \label{e1}
\Gamma^{Z,\gamma}_\mu &=& -ie_0 \ g^{Z,\gamma} \ \left[ \gamma_\mu V_q^{Z,
\gamma}-\gamma_\mu \gamma_5 A_q^{Z,\gamma} + \frac{1}{2m_q} P_\mu S_q^{Z,
\gamma} \right],
\end{eqnarray}
\nn where $e_0$ is the electric charge of the proton, $P=p_1-p_2$ with $p_1,
p_2$ the momenta of the quark and anti--quark and $g^\gamma=1 ,~g^Z= 1/(4s_W
c_W)$ with $s_W^2=1-c_W^2=\sin^2\theta_W$. Because CP is conserved by strong
interactions, terms proportional to $P_\mu \gamma_5$ should be absent and this
fact provides a good check of the calculation. In principle one can also have
scalar and pseudoscalar couplings, $q_\mu$ and $q_\mu \gamma_5$ where
$q=p_1+p_2$ is the momentum of the gauge boson; but in $e^+ e^-$ collisions
these terms give contributions which are proportional to the electron mass and
are therefore totally negligible. At the tree level, $S_q^{Z,\gamma}$ vanish,
while the vector and axial-vector couplings take the usual form:
\begin{eqnarray} \label{e2}
(V_q^Z)^0 \equiv v_q = 2I_q^{3L} -4s_W^2e_q \ \ ,\hspace{0.5cm}
(A_q^Z)^0 \equiv a_q = 2I_q^{3L} \ \ , \hspace{0.5cm}
(V_q^\gamma)^0 =e_q \ \ , \hspace{0.5cm} (A_q^\gamma)^0 =0,
\end{eqnarray}

\nn with $I_q^{3L} = \pm 1/2$ the weak isospin and $e_q$ the electric charge
of the quark. When loop corrections are included, $S_q^{Z,\gamma}$
terms appear and the bare vector and axial-vector couplings are shifted by an
amount
\begin{eqnarray} \label{e3}
\delta V_q^{Z,\gamma}= V_q^{Z,\gamma}-(V_q^{Z,\gamma})^0 \ \ \ , \hspace*{1cm}
\delta A_q^{Z,\gamma}= A_q^{Z,\gamma}-(A_q^{Z,\gamma})^0.
\end{eqnarray}
In previous work \cite{R9,R10} only the corrections to $V_q^Z$ and $A_q^Z$
were considered. We find that even for heavy (top) quarks the corrections
coming from the scalar form factors $S_q^{Z,\gamma}$ are indeed somewhat less
important than the corrections to the couplings that are already present at
tree level.

\vspace*{5mm}
\nn Since we are interested in radiative corrections involving strong
interactions, we only need to consider diagrams involving squark and gluino
loops. As stated in the Introduction, we will include effects proportional to
the mass of the produced quarks. As well known \cite{10}, the supersymmetric
partners of left-- and right--handed massive quarks mix; the mass eigenstates
$\tilde{q}_1$ and $\tilde{q}_2$ being related to the current eigenstates
$\tilde{q}_L$ and $\tilde{q}_R$ by
\begin{eqnarray}\label{e3p}
\tilde{q}_1=\tilde{q}_L \cos \heta +\tilde{q}_R \sin \heta \ \ , \hspace*{1cm}
\tilde{q}_2=- \tilde{q}_L \sin \heta +\tilde{q}_R \cos \heta.
\end{eqnarray}

\nn The mixing angle $\heta$ as well as the masses \mso, \mst\ of the physical
squarks can be calculated from the following mass matrices\footnote{We ignore
generation mixing between squarks, which in case of the MSSM is only induced
radiatively by weak interactions.}:

\ben \label{e4} \beq
{\cal M}^2_{\tilde t} &= \mbox{$ \left( \begin{array}{cc}
m^2_{\tilde{t}_L} + m_t^2 + 0.35 D_Z & - m_t (A_t + \mu \cot \! \beta) \\
- m_t (A_t + \mu \cot \! \beta ) & m^2_{\tilde{t}_R} + m_t^2 + 0.16 D_Z
\end{array} \right) $}; \label{e4a} \\
{\cal M}^2_{\tilde b} &= \mbox{$ \left( \begin{array}{cc}
m^2_{\tilde{t}_L} + m_b^2 - 0.42 D_Z & - m_b (A_b + \mu \tanb) \\
- m_b (A_b + \mu \tanb) & m^2_{\tilde{b}_R} + m_b^2 - 0.08 D_Z
\end{array} \right) $}, \label{e4b} \eeq \een
where $D_Z = M_Z^2 \cos \! 2 \beta$, $\tan \beta$ being the ratio of the
vacuum expectation values of the two neutral Higgs fields of the MSSM
\cite{1}. $m_{\tilde{t}_L,\tilde{t}_R,\tilde{b}_R}$ are soft breaking
masses, $A_{b,t}$ are parameters describing the strength of nonsupersymmetric
trilinear scalar interactions, and $\mu$ is the supersymmetric Higgs(ino)
mass, which also enters trilinear scalar vertices. Notice that the
off--diagonal elements of these squark mass matrices are proportional to the
quark mass. In the case of the supersymmetric partners of the light quarks
mixing between the current eigenstates can therefore be neglected. However,
mixing between $\tilde{t}$ squarks can be sizable and allows one of the mass
eigenstates to be much lighter than the top quark. Sbottom mixing can also be
significant if $\tanb \gg 1$; even in supergravity models with radiative
symmetry breaking \tanb\ can be as large as $m_t/m_b$ \cite{11}. \\

\nn The interactions of the photon and the $Z$ boson with squark current
eigenstates are described by the following lagrangian \cite{1}:
\begin{eqnarray} \label{e5}
{\cal L}_{\tilde{q}\tilde{q}V} = -ie A^\mu \sum_{i=L,R}
e_{q_i} \tilde{q}^*_i \stackrel{\leftrightarrow}{\partial_\mu} \tilde{q}_i \
-\frac{ie}{s_Wc_W} Z_\mu \sum_{i=L,R} (I^{3i}_q -2 e_{q_i}
s_W^2) \tilde{q}^*_i \stackrel{\leftrightarrow}{\partial_\mu} \tilde{q}_i.
\end{eqnarray}

\nn After the introduction of nontrivial squark mixing, this becomes \cite{12}:
\begin{eqnarray} \label{e6}
{\cal L}_{\tilde{q} \tilde{q} V} & = &  -ie A^\mu e_q \left[ \tilde{q}^*_1
\stackrel{\leftrightarrow}{\partial_\mu} \tilde{q}_1 + \tilde{q}^*_2 \stackrel{
\leftrightarrow}{\partial_\mu} \tilde{q}_2 \right] \ - \frac{ie}{s_Wc_W} Z^\mu
\left[ -I^{3L}_q \sin \heta \cos \heta (\tilde{q}^*_1
\stackrel{\leftrightarrow}
{\partial_\mu} \tilde{q}_2 + \tilde{q}^*_2 \stackrel{\leftrightarrow}
{\partial_\mu} \tilde{q}_1 ) \right. \non \\
& & \left. \hspace*{0.5cm} + (I^{3L}_q \cos^2 \heta - s_W^2 e_q ) \tilde{q}^*_1
\stackrel{\leftrightarrow}{\partial_\mu} \tilde{q}_1 +(I^{3L}_q \sin^2 \heta
-s_W^2 e_q ) \tilde{q}^*_2 \stackrel{\leftrightarrow}{\partial_\mu} \tilde{q}_2
)\right].
\end{eqnarray}

\nn Finally, the squark--quark--gluino interaction lagrangian in the presence
of
squark mixing is given by
\begin{eqnarray} \label{e7}
{\cal L}_{\tilde{g}\tilde{q}q} = -i\sqrt{2}g_s T^a \overline{q} \left[ \left(
\cos\heta \tilde{q_1} - \sin \heta \tilde{q_2} \right) \frac{1+\gamma_5}{2} -
\left( \sin \heta \tilde{q_1} + \cos  \heta \tilde{q_2}
\right)\frac{1-\gamma_5}
{2} \right] \tilde{g}^a + {\rm h.c.},
\end{eqnarray}

\nn where $g_s$ is the strong coupling constant and $T^a$ are SU(3)$_C$
generators. Note that in eq.~(\ref{e7}) we have assumed $M_{\tilde{g}}>0$. \\

\vspace*{-2mm}

\nn Including the corrections due to the squark/gluino vertex diagram shown
in Fig.~1a, and taking into account the mixing between the left and
right--handed squarks as well as the finite mass of the external quarks, the
photon couplings to quarks are shifted by:

\ben \label{e8} \beq
(\delta V_q^\gamma)_{\rm VE} &= \frac{4}{3}\frac{\alpha_s}{\pi} \frac{e_q }{2}
\left[ C_3^{11} +C_3^{22} \right]; \label{e8a} \\
(\delta A_q^\gamma)_{\rm VE} &= \frac{4}{3}\frac{\alpha_s}{\pi} \ \frac{e_q}
{2} \ \cos2 \heta \ \left[ C_3^{11} -C_3^{22} \right]; \label{e8b} \\
(S_q^\gamma)_{\rm VE} &= - \frac{4}{3}\frac{\alpha_s}{\pi} \frac{e_q }{2}
\left[ \mt^2
(C_2^{11}+C_2^{22} )- \mt \mg\sin 2 \heta (C_1^{11}- C_1^{22}) \right].
\label{e8c} \eeq \een
\nn The corresponding shifts of the $Z$ boson couplings to quarks are:
\ben \label{e9} \beq
(\delta V_q^Z)_{\rm VE} &= \frac{4}{3} \frac{\alpha_s}{\pi} \left[
(\is \cos^2 \heta -2s_W^2 e_q) C_3^{11} +(\is \sin^2 \heta-2s_W^2e_q)
C_3^{22} \right];  \label{e9a} \\
(\delta A_q^Z)_{\rm VE} &= \frac{4}{3}\frac{\alpha_s}{\pi} \left[ (\is
\cos^2 \heta -2s_W^2 e_q) \cos2 \heta C_3^{11} -(\is \sin^2 \heta-
2s_W^2e_q)\cos2 \heta C_3^{22} \right. \non \\
&  \hspace*{1.4cm} \left. +  \ I_q^{3L} \sin^2 2 \heta (C_3^{12}+C_3^{21}
) \right]; \label{e9b} \\
(S_q^Z)_{\rm VE} &= - \frac{4}{3}\frac{\alpha_s}{\pi} \left[ (\is
\cos^2 \heta -2s_W^2 e_q) (\mt^2 C_2^{11}- \mt \mg\sin 2 \heta C_1^{11})
+(\is \sin^2 \heta -2s_W^2e_q) \right. \non \\
&  \hspace*{1.9cm} \left. \times (\mt^2 C_2^{22}+ \mt \mg\sin 2
\heta C_1^{22}) + I_q^{3L} \sin2 \heta \cos2 \heta \mt \mg (C_1^{12}+C_1^{21})
\right]. \label{e9c} \eeq \een

\nn For $i,j=1,2$ the $C_{1,2,3}^{ij}$ are defined as $C_k^{ij} \equiv
C_k (s, m_q, \msi , \msj, \mg)$. We use the Passarino--Veltman reduction to
scalar
integrals \cite{R5}, and the intermediate function $C^{ij}_+$ and $C^{ij}_-$
defined as (note that we use a slightly different notation than in \cite{R6}):
\beq \label{e10}
C^{ij}_+ &= \frac{-1}{2s\beta_q^2} \left[ 2B_0(s,\msi,\msj)-B_0(\mt^2,\mg,
\msi) - B_0(\mt^2,\mg,\msj) \right. \non \\ & \hspace*{1.5cm} \left.
+(2\mg^2+2\mt^2-\msi^2 -\msj^2) C^{ij}_0
\right] \non \\
C^{ij}_- &= \frac{1}{2s} \left[ B_0(\mt^2,\mg,\msi)-B_0(\mt^2,\mg,\msj)+
(\msj^2 -\msi^2) C^{ij}_0 \right]. \eeq

\nn In terms of these functions, the $C_k^{ij}$ are given by:
\ben \label{e11} \beq
C_1^{ij}&= C^{ij}_0 - 2C_+^{ij}, \label{e11a} \\
C_2^{ij}&= 2C_+^{ij}-\frac{4}{s\beta_q^2} \left[ C^{ij}_3 + \frac{1}{2}
(\msi^2 +\msj^2-2\mg^2 -2\mt^2) C_+^{ij} - \frac{1}{2}B_0(s,\msi,\msj)
\right. \non \\
& \hspace*{2.6cm} \left. + \frac{1}{4} B_1 (\mt^2, \mg, \msi)+
\frac{1}{4} B_1(\mt^2,\mg, \msj)  \right], \label{e11b} \\
C_3^{ij}&= \frac{1}{4} \left[ 2 \mg^2 C^{ij}_0 +1+B_0(s,\msi,\msj) +
(\msi^2 +\msj^2-2\mt^2 -2\mg^2) C^{ij}_+ \right. \non \\
& \hspace*{.8cm} \left. +(\msi^2 -\msj^2)C^{ij}_- \right].
\label{e11c} \eeq \een

\nn Here, $B_1$ is given by
\begin{eqnarray} \label{e12}
B_1(s,m_1,m_2) = \frac{1}{2s} \left[(s+m_1^2-m_2^2)B_0(s,m_1,m_2)+A_0(m_2)
-A_0(m_1) \right],
\end{eqnarray}
and the functions $A_0, B_0$ and $C_0$ correspond to the scalar one, two and
three point functions \cite{R7}, respectively, and are given in Appendix. \\

\nn The renormalized vertices are derived by adding the
counterterm originating from the on--shell self--energies of the external
quarks, Fig.~1b. Following the procedure outlined in \cite{R6,R8}, one obtains
\begin{eqnarray} \label{e13}
(\delta V_q^\gamma)_{\rm CT}  = e_q \delta Z_V \hspace*{1cm} &,& \hspace{0.5cm}
(\delta A_q^\gamma)_{\rm CT}  = e_q \delta Z_A, \hspace{1cm} \non \\
(\delta V_q^Z)_{\rm CT}  = v_q \delta Z_V + a_q \delta Z_A \hspace{0.5cm} &,&
\hspace{0.5cm} (\delta A_q^Z)_{\rm CT}  = a_q \delta Z_V +v_q \delta Z_A.
\end{eqnarray}
\nn Here, $\delta Z_V$ and $\delta Z_A$ are given by
\ben \label{e14} \beq
\delta Z_V &= -\frac{1}{3} \frac{\alpha_s}{\pi}\left[ B_1(m_q^2,\mg,\mso)
+B_1 (m_q^2,\mg,\mst)+2m_q^2((B_1'(m_q^2, \mg,\mso )+B_1'(m_q^2,\mg,\mst))
\right. \non \\
& \left. \hspace*{1.6cm} -2m_q \mg \sin 2 \tilde{\theta} (B_0'(m_q^2, \mg,\mso)
-B_0'(m_q^2, \mg,\mst)) \right] \label{e14a} \\
\delta Z_A &= -\frac{1}{3} \frac{\alpha_s}{\pi} \cos 2 \tilde{\theta} \left[
B_1(m_q^2,\mg,\mso) - B_1 (m_q^2, \mg,\mst) \right]. \label{e14b}
\eeq \een

\nn The full SUSY--QCD correction to the vectorial and axial couplings is just
the sum of the unrenormalized vertex correction and the quark self-energy
counterterms\footnote{Note that in this convention the vector couplings $V_q$
and scalar couplings $S_q$ do not reduce to their SM values even in the limit
of
infinite squark masses; of course, very heavy squarks and gluinos do decouple
from physical observables such as cross sections. Physically equivalent
results can be obtained by ignoring the diagrams of Fig.~1b, and performing
the renormalization by simply substracting the corrections at zero momentum
transfer, $s=0$; in this scheme separate counterterms for $V_q$ and $S_q$ can
be defined, so that each coupling by itself reduces to its SM value in the
limit of large sparticle masses.}:
\ben \label{e15} \beq
\delta V_q^{\gamma,Z} &=(\delta V_q^{\gamma , Z})_{\rm VE}  +
(\delta V_q^{\gamma , Z})_{\rm CT},  \label{e15a} \\
\delta A_q^{\gamma,Z} &=(\delta A_q^{\gamma , Z} )_{\rm VE}  +
(\delta A_q^{\gamma , Z})_{\rm CT}. \label{e15b}
\eeq \een

\nn The expressions (\ref{e8}) -- (\ref{e14}) are rather cumbersome. For many
applications squark mixing can be neglected. If one in addition assumes
approximate degeneracy for the squarks, $\mso=\mst\equiv\msq$, the corrections
simplify considerably, and one finds:
\beq \label{e16}
\delta V_q^{\gamma,Z} &=  \frac{4}{3}\frac{\alpha_s}{\pi} (V_q^{\gamma,Z})^0 C
\hspace*{0.5cm} , \hspace*{0.5cm}
\delta A_q^{\gamma,Z} = \frac{4}{3}\frac{\alpha_s}{\pi} (A_q^{\gamma,Z})^0 C,
\non \\
S_q^{\gamma,Z} &= - \frac{4}{3}\frac{\alpha_s}{\pi} m_q^2 (V_q^{\gamma,Z})^0
C_2(s,m_q,\msq,\msq,\mg)
\eeq
where $C$ is given by:
\begin{eqnarray} \label{e17}
C \equiv C_3(s,m_q,\msq,\msq,\mg)-\frac{1}{2}B_1(m_q^2,\msq,\mg)
- \frac{1}{2} m_q^2 B'_1(m_q^2,\msq,\mg).
\end{eqnarray}

\nn Furthermore, for massless final state quarks
the $S_q$ term vanishes and the correction to the axial and vector couplings
can be expressed by a single two--dimensional integral as:
\begin{eqnarray} \label{e18}
C \simeq C_3-\frac{1}{2} B_1 = \frac{1}{2} \int_0^1 x{\rm d}x \int_0^1
{\rm d}y \log \frac{ x(\msq^2-\mg^2)+\mg^2}{-sx^2y
(1-y)+x(\msq^2-\mg^2)+\mg^2},
\end{eqnarray}

\nn in agreement with \cite{R9}. For large squark and gluino masses,
$\msq ,\mg \gg s$, the correction is just
\begin{eqnarray} \label{e19}
C \rightarrow \frac{s}{12} \frac{1}{(\msq^2-\mg^2) ^4}& & \left[ \frac{1}{3}
(\msq^2-\mg^2)^3-\frac{1}{2}\mg^2 (\msq^2-\mg^2)^2 + \mg^4(\msq^2-\mg^2)
\right. \non \\ & & \left.
-\mg^6 \log \frac{\msq^2}{\mg^2} \right].
\end{eqnarray}
\nn If in addition the gluino mass can be neglected compared to the squark
mass, one simply obtains $C \simeq s/(36\msq^2)$. \\

\nn In terms of the vertices (\ref{e1}), the differential cross section
$d \sigma (e^+e^- \rightarrow \qqbar)/ d \cos \theta$ reads (we define
$\theta$ as the angle between the quark and the incoming positron):
\begin{eqnarray} \label{e20}
\frac{{\rm d}\sigma}{ {\rm d}\cos \theta} &=& \frac{3}{8} N_c \beta_q \left\{
D_{\gamma \gamma} e_e^2 \left[ (2- \beta_q^2 \st)(V_q^\gamma)^2 +\bq^2 (1+\ct)
(A^\gamma_q)^2 -2 \beta_q^2 \st V_q^\gamma S_q^\gamma \right] \right. \non \\
&+& D_{Z \gamma} e_e v_e \left[ (2-\beta_q^2 \st) V_q^\gamma V_q^Z +
\bq^2 (1+\ct) A^\gamma_q A_q^Z - \beta_q^2 \st (V_q^\gamma S_q^Z+S_q^\gamma
V_q^Z) \right] \non \\
&+& D_{ZZ} (v_e^2+a_e^2) \left[ (2-\beta_q^2 \st)(V_q^Z)^2 +\bq^2
(1+\ct) (A_q^Z)^2 -2 \beta_q^2 \st V_q^Z S_q^Z \right] \non \\
&+& \left. 2 D_{Z \gamma} e_e a_e \bq \cos \theta (V_q^\gamma A_q^Z +V_q^Z
A_q^\gamma) + 8D_{ZZ} a_e v_e \beta_q \cos \theta V_q^Z A_q^Z \right \}.
\end{eqnarray}

\nn Here $N_c=3$ is the color factor, and $\beta_q=(1-4m_q^2/s)^{1/2}$ the
velocity of the final quarks. In eq.~(\ref{e20}) the leading
electroweak radiative corrections have been included by introducing the
quantities $D_{\alpha\beta}, \ \alpha,\beta \ = \ \gamma, Z$, which are
defined in terms of the Fermi coupling constant $G_F$ and the running QED
coupling $\alpha(s)$:
\begin{eqnarray} \label{e21}
D_{\gamma \gamma}= \frac{4 \pi \alpha^2 (s)}{3s} \ \ \ , \ \ \
D_{ZZ }= \frac{G_F^2}{96\pi} \frac{M_Z^4s}{(s-M_Z^2)^2 +(s\Gamma_Z/M_Z)^2 }
\ \ \ , \non \\
D_{Z \gamma}= \frac{G_F \alpha(s)}{3\sqrt{2}} \frac{M_Z^2(s-M_Z^2)}{(s-M_Z^2)^2
+(s\Gamma_Z/M_Z)^2 }. \hspace*{2cm}
\end{eqnarray}

\vspace*{5mm}
\nn At ${\cal O}(\alpha_s)$, the deviations of the total cross section and the
forward--backward asymmetry from the tree level values, $\delta \sigma =\sigma
-\sigma^0$ and $\delta A_{FB} =A_{FB}-A_{FB}^0$, are then
\ben \label{e22} \beq
\delta \sigma =& N_c \beta_q \left\{ D_{\gamma \gamma}e_e^2 \left[(3-\beta_q^2)
e_q \delta V_q^\gamma - \beta_q^2 e_q S_q^\gamma \right] \right. \non \\
& \hspace*{.8cm} \left. + D_{ZZ} (v_e^2+a_e^2)
\left[ (3-\beta_q^2)v_q \delta V_q^Z + 2 \beta_q^2 a_q \delta A_q^Z
- \beta_q^2 v_q S_q^Z \right]  \right. \non \\ & \hspace*{.8cm} \left.
+ D_{Z \gamma} e_e v_e \left[
\frac{3-\beta_q^2}{2} (v_q \delta V_q^\gamma + e_q \delta V_q^Z)+ \beta_q^2
a_q \delta A_q^\gamma -\frac{\beta_q^2}{2} (e_qS_q^Z + v_q S_q^\gamma) \right]
\right\}, \label{e22a} \\
\delta A_{FB} =& \frac{ D_{Z \gamma} e_e a_e (e_q \delta A_q^Z +a_q \delta
V_q^\gamma+v_q \delta A_q^\gamma) + 4 D_{ZZ}a_e v_e (\delta V_q^Z a_q+\delta
A_q^Z v_q) } { D_{Z \gamma} e_e a_e e_q a_q + 4 D_{ZZ} a_e v_e a_q v_q}
\ - \ \frac{ \delta \sigma} {\sigma^0}. \label{e22b}
\eeq \een

\nn These expressions have to be supplemented by including the standard QCD
corrections; the formulae for the cross section and the forward--backward
asymmetry in the massive case can be found in \cite{R1}. In the case of the
cross section, one can however use the Schwinger formulae \cite{R2}, which
provide a very good approximation to the exact result; this is done by
performing the following substitution $(\alpha, \beta =\gamma, Z$):
\ben \label{e23} \beq
(V_q^\alpha)^0 (V_q^\beta)^0 \longrightarrow & (V_q^\alpha)^0 (V_q^\beta)^0
\left\{ 1 +\frac{4}{3} \frac{\alpha_s}{\pi} \left[ \frac{\pi^2 }{2\beta_q}-
\frac{3+\beta^2_q}{4} \left( \frac{\pi^2}{2}-\frac{3}{4} \right) \right]
\right\}, \label{e23a} \\
(A_q^\alpha)^0 (A_q^\beta)^0 \longrightarrow & (A_q^\alpha)^0 (A_q^\beta)^0
\left\{ 1 +\frac{4}{3} \frac{\alpha_s}{\pi} \left[ \frac{\pi^2}{2\beta_q}-
\left(\frac{19}{10}-\frac{22}{5}\beta_q+\frac{7}{2}\beta_q^2 \right) \left(
\frac{\pi^2} {2}-\frac{3}{4} \right) \right] \right\}. \label{e23b}
\eeq \een
\nn A similarly simple yet very accurate substitution also exists for
standard QCD corrections to the forward--backward asymmetry \cite{R3}:
\begin{eqnarray} \label{e24}
(V_q^\alpha)^0 (A_q^\beta)^0 & \longrightarrow & (V_q^\alpha)^0 (A_q^\beta)^0
\left\{ 1 + \frac{\alpha_s}{\pi} \ \frac{2}{\beta_q} (3-\beta^2_q) \sqrt{1-
\beta_q^2}  \ \right\}
\end{eqnarray}

\vspace*{5mm}
\nn Finally, on top of the $Z$ resonance these expressions simplify
considerably. Besides the fact that only the $Z$ exchange contribution has to
be taken into account, one can neglect to a good approximation the quark
masses  (except possibly in the $\tilde{b}$ mass matrix; see below) since top
decays of the $Z$ boson are kinematically forbidden (to achieve a better
precision one can eventually include the leading mass effects in the Born term
as well as the QCD corrections in the case of the bottom quark; see
\cite{R4}). In this case, the $S_q$ terms vanish and the deviation of the
decay width $\Gamma_q =\Gamma(Z \rightarrow q\bar{q})$ and the
forward--backward asymmetry $A_{FB}$ from their tree level values are simply
given by
\begin{eqnarray}
\frac{\delta \Gamma_q } {\Gamma^0_q} & = & 2 \frac{v_q \delta V_q^Z +a_q
\delta A_q^Z}{v_q^2+a_q^2} \non \\
\frac{ \delta A_{FB}}{A^0_{FB}} &= & \frac{v_q \delta A_q^Z +a_q \delta V_q^Z}
{a_q v_q} - 2 \frac{v_q \delta V_q^Z +a_q \delta A_q^Z}{v_q^2+a_q^2}
\end{eqnarray}

\section*{3. Results}
\nn We are now in a position to present some numerical examples. In Fig.~2 we
show SUSY QCD corrections to the hadronic decay width of the Z boson; the
solid (dashed) curves are for \bbbar\ (\ccbar) final states. We have set the
$A$ parameters in the squark mass matrices (\ref{e4}) to zero, and have
assumed equal SUSY breaking masses for all squarks, denoted by \msqav.
Moreover, in this figure we have assumed that all parameters entering the
squark mass matrices, as well as the gluino mass, can be varied independently
(``global SUSY'' scenario).
The four upper curves are for negligible mixing between $L$ and $R$ squarks.
Even in this case the ``D--terms'' ($D_Z$ in eqs.(\ref{e4})) lead to
nonnegligible mass splitting between squarks of different flavour, if
$\tanb \ne 1$. In particular, for $\tanb > 1$ (which is favoured by
supergravity models \cite{3}), $\tilde{u}$ type squarks are lighter than
$\tilde{d}$ type squarks; as a result, for a given value of \msqav\ the
corrections to \ccbar\ production are larger than those to \bbbar\
production.\\

\nn The uppermost curves in Fig.~2 have been obtained by chosing a very small
gluino mass, $\mg=3$ GeV. A gluino of this mass could have escaped all
experimental searches, provided squarks are heavier than 100 GeV or so
\cite{13}. Although a careful study showed \cite{14} that a GeV gluino does not
reduce the slight discrepancy between values of \as\ extracted from low energy
experiments and those derived from event shape variables measured at $\rs
\simeq M_Z$, present measurements cannot exclude its existence, either. It
should also be noted that squark mass bounds from hadron colliders \cite{8}
might be invalidated by the presence of such a light gluino. This is
because in this scenario, squarks predominantly decay into gluinos, which
lose a considerable fraction of their energy in QCD radiation prior to their
decay, thereby leading to a rather soft missing $p_T$ spectrum \cite{15}.
{}From Fig.~2 we conclude that 1--loop SUSY QCD corrections to the hadronic
width of the $Z$ boson could amount to about $0.3 \%$, or about $8 \%$ of the
standard QCD correction. For very light gluinos, 2--loop SUSY QCD corrections
are also not entirely negligible \cite{R9}; they amount to about $-2\%$ of
the standard QCD corrections \cite{14}. Altogether SUSY QCD corrections to
$\Gamma_{\rm had}$ therefore amount to at most +6\% of the standard QCD
corrections, for a gluino mass of a few GeV and squark masses around 100 GeV.
In this scenario the value of \as\ extracted from the measurement of
$\Gamma_{\rm had}$ would therefore have to be reduced by about 6\%. At the
same time, in the presence of light gluinos the value of \as\ derived from
event shape variables has to be increased by about 8\% \cite{14}. The net
result is that the present small discrepancy between these two determinations
of \as\ is diminished\footnote{Strictly speaking the analysis of
Ref.~\cite{14} is valid only for very heavy squarks; however, we expect it to
hold also for squark masses around 100 GeV, since squark exchange diagrams
contributing to $\qqbar \tilde{g} \tilde{g}$ production are not enhanced by
large logarithms, unlike diagrams where a gluino pair is produced from a
gluon.\\}. \\

\nn If we chose gluino and squark masses above the region excluded by
hadron collider searches \cite{8} the maximal size of the corrections to
$Z$ partial widths drops by about a factor of 2, as illustrated by the
curves for $\mu=0$ and \mg\ = 160 GeV in Fig.~2. Moreover, squark mass
splitting due to D--terms becomes less important, so that to good
approximation the simplified expressions (\ref{e16}) and (\ref{e19}) can be
used.\\

\nn Finally, the lowest curve in Fig.~2 demonstrates that squark mixing can
have sizable effects already for $\tilde{b}$ squarks. The off--diagonal
elements of the $\tilde{b}$ squark mass matrix (\ref{e4b}) can be substantial
if $\tanb \gg 1$ and $\mu$ is not too small. Indeed, for the parameters chosen
in Fig.~2, the lighter $\tilde{b}$ eigenstate would be lighter than 45 GeV, in
violation of LEP bounds \cite{7}, unless $\msqav\geq 180$ GeV\footnote{Note
that hadron collider data do not exclude the existence of a single light
squark species, if the mass of the lightest neutralino exceeds about 15 GeV
\cite{15a}.}. In this scenario the correction to the partial width into \bbbar\
pairs is negative. The corrections to the $u, d, s, c$ partial widths for the
same set of parameters are still positive, however, leading to a very small
correction to the total hadronic width of the $Z$ boson. In order to test this
scenario experimentally one would thus have to measure the \bbbar\ cross
section with a precision of a fraction of 1\%, which appears to be quite
difficult. \\

\nn In Fig.~3 we plot the correction to the total cross section for the
production of light quarks at \rs = 500 GeV. In this figure we have switched
off both squark mixing (by setting $A_q = \mu = 0$) and squark mass splitting
through D--terms (by setting \tanb=1); however, the previous figure showed
that results for nonzero $\mu$ and $\tanb \ne 1$ are quite similar unless
the gluino is very light or \tanb\ is very large. We see that the corrections
reach a maximum at $\msq \simeq 0.4 \rs \simeq 200$ GeV, almost independently
of the value of \mg. If both \msq\ and \mg\ are much smaller than \rs\ the
corrections become negative; in the limit of exact SUSY ($\msq \rightarrow m_q,
\ \mg \rightarrow 0$) one encounters logarithmic infrared divergencies.
For $\msq > 200$ GeV the size of the corrections decreases rapidly. However,
even if squarks are not accessible to the accelerator we study, i.e. for
$\msq > \rs/2$, the correction can be as large as $+1\%$, or about one third
the
standard QCD correction. We also observe that the correction depends less
sensitively on the gluino mass than on the squark mass; this has also been
found in Ref.~\cite{R9}. \\

\nn In figs. 4a,b we present results for SUSY QCD corrections to \ttbar\
production at \rs\ = 500 GeV, for $m_t$ = 150 GeV. We have fixed the gluino
mass to 250 GeV and chosen $\tanb=2$. The dashed curves are again valid for
a ``global SUSY'' model, with $m_{\tilde{t}_L} = m_{\tilde{t}_R} = A_t$ and
$\mu = 500$ GeV. In contrast, the solid curves are for a supergravity
scenario, where scalar masses are assumed to be equal to each other, and also
equal to the scalar trilinear interaction parameters $A_q$, at the scale
of Grand Unification, $M_X \simeq 10^{16}$ GeV. The parameters at the weak
scale have then be computed by solving a set of coupled renormalization
group equations \cite{4}; for simplicity we have treated them using the
analytical approximations given in Ref.~\cite{16}. Notice that in this model
$\mu$ is no longer a free parameter, but determined by the requirement of
correct SU(2)$\times$U(1) symmetry breaking, $M_Z = 91.1$ GeV. Moreover,
$m_{\tilde{t}_R}$ is considerably smaller than $m_{\tilde{t}_L}$ at the
weak scale, due to quantum corrections involving the $t$ quark Yukawa
coupling. \\

\nn The results of Fig.~4 are presented as a function of the mass of the
lighter $\tilde{t}$ eigenstate. The mass of the heavier eigenstate varies
between 390 and 620 GeV in the global SUSY model, and between 750 and 1100
GeV in the SUGRA scenario we are considering. Moreover, in the former case
the $\tilde{t}$ mixing angle is close to $45^\circ$, since the diagonal
elements of the $\tilde{t}$ mass matrix (\ref{e4a}) are almost equal, while
in the latter case the angle is considerably larger than $45^\circ$, so that
the light eigenstate is predominantly $\tilde{t}_R$. We see that the
correction to the total cross section, shown in Fig.~4a, is not very
sensitive to the differences between the two models we are studying. The
reason is that the total cross section is dominated by the photon exchange
contribution, which does not depend on $\tilde{t}$ mixing, see
eqs.(\ref{e8a}) and (\ref{e22a}). The corrections are smaller than for the
production of light quarks (with $\msq = m_{\tilde{t}_1}$) since in case of
\ttbar\ production practically only one squark contributes in the loop,
the heavier $\tilde{t}$ eigenstate being much more massive.

\vspace{5mm}
\nn In contrast, the forward--backward asymmetry (\ref{e22b}) is sensitive
to the $Z$ exchange contribution, and hence to $\tilde{t}$ mixing; Fig.~4b
shows that for small $m_{\tilde{t}_1}$ even the sign of the correction
differs for the two models. Unfortunately the absolute value of this
correction is always less than 0.5\%; one would probably need a dedicated
``top factory'' to achieve this level of precision. SUSY QCD corrections to
the forward--backward asymmetries of light quarks are always well below
0.1\%, and can therefore safely be neglected.

\section*{4. Summary and Conclusions}
\nn In this paper we have presented explicit expressions for the $\gamma
\qqbar$ and $Z \qqbar$ vertices, allowing for mixing between the
superpartners of left-- and right--handed quarks as well as for unequal
squark masses in the loop. We have found corrections to the total cross
section (or, on the $Z$ pole, to the hadronic decay width of the $Z$) to be
usually positive, unless both  squark and gluino masses are much smaller
than the centre--of--mass energy \rs. In the limit of no squark mixing and
degenerate squark masses we reproduce the results of Ref.~\cite{R9}.
For massless quarks,
corrections are largest if $\msq \simeq 0.4 \rs$, i.e. just above the threshold
for open squark production, where they can reach $+2\%$; they fall below
$1\%$ at $\msq \simeq 0.6 \rs$, the exact value depending on the gluino mass.
If the gluino mass is just a few GeV and squark masses are around 100 GeV,
which still appears to be allowed experimentally, supersymmetric QCD
corrections might help to improve the agreement between values of \as\
derived from event shape variables at $\rs \simeq M_Z$ and from the total
hadronic decay width of the $Z$ boson. We also found that $\tilde{b}$ mixing
can be important, and can even flip the sign of the correction. \\

\nn The corrections to the total \ttbar\ production cross section are usually
smaller than for the case of light squarks, for a given mass of the lightest
squark eigenstate of a given flavor. The reason is that $\tilde{t}$ mixing
pushes the mass of the heavier $\tilde{t}$ eigenstate to such large values
that its contribution is essentially negligible. We also computed
corrections to the forward--backward asymmetry, and found them to be well
below $0.1\%$ for light quarks. In case of $t$ quarks these corrections are
sensitive to the details of $\tilde{t}$ mixing, unlike the total \ttbar\
cross section; however, even for $t$ quarks the forward--backward asymmetry is
changed by less than $0.5\%$. \\

\nn We conclude that, barring the existence of a very light gluino,
supersymmetric QCD corrections to the production of \qqbar\ pairs in \eplem\
annihilation are probably only observable at energies above the open squark
threshold. They will therefore not be useful as a tool to search for
supersymmetry; however, after the discovery of a ``new physics'' signal they
might allow to confirm its interpretation in terms of supersymmetry.

\vspace*{.5cm}
\subsection*{Acknowledgements}
One of us (A.D.) thanks Vernon Barger and the members of the Phenomenology
group at the Physics Department of the University of Wisconsin for providing
a hospitable working environment during his stay. This work was supported in
part by the U.S. Department of Energy under contract No. DE-AC02-76ER00881,
and in part by the Wisconsin Research Committee with funds granted by the
Wisconsin Alumni Research Foundation. The work of M.D. was supported by
a grant from the Deutsche Forschungsgemeinschaft under the Heisenberg
program.

\renewcommand{\theequation}{A.\arabic{equation}}
\setcounter{equation}{0}

\section*{Appendix: Scalar Loop Integrals}

\nn In this Appendix we collect expressions that allow to evaluate the
loop functions that appear in sec. 2.
The scalar one, two and three point functions, $A_0, B_0$ and $C_0$
are defined as \cite{R8}:
\begin{eqnarray} \label{a1}
A_0(m_0) &=& \frac{(2\pi\mu)^{n-4}}{i\pi^2} \int \ \frac{d^nk}{k^2-m_0^2+
i\epsilon}, \non \\
B_0(s,m_1,m_2) &=& \frac{(2\pi\mu)^{n-4}}{i\pi^2} \ \int \frac{d^nk}{
(k^2-m_1^2 + i\epsilon) [(k-q)^2-m_2^2+i \epsilon] }, \\
C_0(s, m_1, m_2, m_3) &=& \frac{(2\pi\mu)^{n-4}}{i\pi^2} \int \frac{d^nk}{
[(k-p_1)^2-m_1^2 +i\epsilon][(k-p_2)^2-m_2^2 +i\epsilon](k^2-m_3^2+i\epsilon)}.
\non
\end{eqnarray}

\nn Here $n$ is the space--time dimension and $\mu$ the renormalisation scale.
\\

\nn After integration over the internal momentum $k$, the function $A_0$ is
given by:
\begin{eqnarray} \label{a2}
A_0(m_0)= m^2_0 \left[ 1+ \Delta_0 \right]  \ \ , \hspace*{1cm}
\Delta_i = \frac{2}{4-n} - \gamma_E + \log (4 \pi) +\log \frac{\mu^2}{m_i^2},
\end{eqnarray}
\nn where $\gamma_E$ is Euler's constant. The function $B_0$ and its
derivative with respect to $s$, $B'_0$, are given by
\begin{eqnarray} \label{a3}
B_0(s,m_1,m_2) &=& \frac{1}{2}(\Delta_{1}+ \Delta_{2}) +2 +\frac{m_1^2-m_2^2}
{2s} \log \frac{m_2^2}{m_1^2} +\frac{x_+ -x_-}{4s} \log \frac{x_-}{x_+}, \non
\\
B_0'(s,m_1,m_2) &=& -\frac{1}{2s} \left[ 2+\frac{m_2^2-m_1^2}{s} \log
\frac{m_1^2} {m_2^2} + \frac{2}{s} \frac{(m_1^2-m_2^2)^2 - s(m_1^2+m_2^2)}
{x_+-x_-} \log\frac{x_-}{x_+} \right],
\end{eqnarray}
\nn with
\begin{eqnarray} \label{a4}
x_\pm =s- m_1^2 -m_2^2 \pm \sqrt{ s^2-2s (m_1^2+m_2^2)+(m_1^2 - m_2^2 )^2}
\end{eqnarray}
\nn Note that the $x_\pm$ can be complex. For $ (m_1 - m_2)^2 < s <
(m_1 + m_2)^2$, the logarithms appearing in eqs.(\ref{a3}) can be
expressed in terms of an $\arctan$ of a real argument. When writing these
equations we have ignored the imaginary parts of $B_0$ and $B_0'$; they are
not relevant for us, since to next--to--leading order we are only interested in
the interference between the (real) tree--level and one--loop apmlitudes.

\vspace*{5mm}
\nn In this paper we need the three point scalar function $C_0$ only for
$p_1^2 = p_2^2 = m_q^2$; in this case it can be written in integral form as
\begin{eqnarray} \label{a5}
C_0(s,m_q,m_1,m_2,m_3) = - \int_0^1 dy \int_0^y dx \left[ ay^2+bx^2+cxy
+dy+ex+f \right]^{-1},
\end{eqnarray}
where
\begin{eqnarray} \label{a6}
a=m_q^2 \ , \ \ b=s \ , \ \ c=-s \ , \ \ d = m_2^2 -m_3^2 - m_q^2 \ ,
\ \ e = m_1^2 -m_2^2 \ , \ \ f=m_3^2 -i \epsilon. \hspace*{3mm}
\end{eqnarray}

\nn $C_0$ can be expressed in terms of a sum of Spence functions
Li$_2(x)=-\int_0^1 dt \log(1-xt)/t$:
\begin{eqnarray} \label{a7}
C_0(s,m_q,m_1,m_2,m_3)= - \frac{1}{s \beta_q} \ \sum_{i=1}^{3} \sum_{j=+,-}
\ (-1)^i \left[ {\rm Li_2} \left( \frac{x_i} {x_i-y_{ij} } \right) -
{\rm Li_2} \left( \frac{x_i-1}{x_i-y_{ij}} \right) \right],
\end{eqnarray}
where we have defined
\begin{eqnarray} \label{a8}
x_1=\frac{2d+e(1-\beta_q)}{2s\beta_q}+\frac{1}{2}(1-\beta_q) \ \ ,
\hspace*{1cm} & & y_{1\pm} =\frac{-c-e \pm \sqrt{(c+e)^2-4b(a+d+f)} } {2b},
\non \\
x_2=\frac{2d+e(1-\beta_q)}{s\beta_q (1+\beta_q)} \hspace*{2.2cm} \ \ ,
\hspace*{1cm}
& & y_{2\pm} =\frac{-d-e \pm \sqrt{(d+e)^2-4f(a+b+c)} } {2(a+b+c)}, \non \\
x_3= -\frac{2d+e(1-\beta_q)}{s\beta_q (1-\beta_q)} \ \  \hspace*{1.9cm} ,
\hspace*{1cm} & & y_{3\pm} =\frac{-d \pm \sqrt{d^2-4af}} {2a}.
\end{eqnarray}

\nn The $x_i$ and $y_{i \pm}$ can again be complex. Eq.~(\ref{a7}) is only
valid above the $\qqbar$ threshold, i.e. for $s > 4 m_q^2$. Below the
threshold analytical continuation of complex logarithms requires the
introduction of additional terms; see Ref.~\cite{R5}.

\newpage
\section*{Figure Captions}

\vspace*{1cm}

\renewcommand{\labelenumi}{Fig. \arabic{enumi}}
\begin{enumerate}
\item  
Vertex (1a) and self--energy (1b) corrections to
$\eplem \rightarrow \qqbar$ from supersymmetric QCD.

\vspace{5mm}
\item  
Supersymmetric QCD corrections to the decay width of the $Z$ boson into
\ccbar\ (dashed) and \bbbar\ pairs (solid). We have assumed a ``global SUSY''
scenario, where all parameters of the squark mass matrices can be varied
independently. For $\tilde{c}$ and $\tilde{b}$ squarks the $A$--terms are
always negligible; the other parameters are as indicated in the figure.
Notice that \msqav\ is the common SUSY breaking diagonal squark mass, which
is also the average first generation squark mass, since D--term contributions
cancel after summing over a complete generation. The lowest curve ends at
\msqav=180 GeV since for even smaller values, $m_{\tilde{b}_1} < 45$ GeV.

\vspace{5mm}
\item  
Supersymmetric QCD corrections to the total cross section for the production
of light \qqbar\ pairs at an \eplem\ collider with \rs=500 GeV. In this figure
we have switched off squark mixing and chosen \tanb=1, so that the
superpartners of all 5 light quarks have the same mass \msq.

\item 
Supersymmetric QCD corrections to the total cross section (a) and
forward--backward asymmetry (b) for \ttbar\ pair production at a 500 GeV
\eplem\ collider, as a function of the mass of the lighter $\tilde{t}$
eigenstate. The dashed curves are for a ``global SUSY'' model with
$\mu=500$ GeV, while the solid curves are for a supergravity scenario with
radiative symmetry breaking, where $\mu$ is a derived quantity, as described
in the text. In the former case we have assumed $m_{\tilde{t}_L} =
m_{\tilde{t}_R} = A_t$ at the weak scale, while in the latter case these
relations are valid only at the Grand Unified scale.

\end{enumerate}
\end{document}